\documentclass[
 reprint,
 superscriptaddress,
 amsmath,amssymb,
 aps,
 pra,
 floatfix,
]{revtex4-2}
\pdfoutput=1

\usepackage{graphicx}
\usepackage{dcolumn}
\usepackage{bm}
\usepackage[colorlinks=true, linkcolor=black, urlcolor=black, citecolor=black]{hyperref}

\newcommand{\dcwidth}{140mm}

\usepackage[normalem]{ulem}

\begin{document}

\preprint{APS/123-QED}

\title{Coulomb-mediated antibunching of an electron pair surfing on sound}

\author{Junliang Wang}
    \affiliation{Universit\'e Grenoble Alpes, CNRS, Grenoble INP, Institut N\'eel, F-38000 Grenoble, France}
\author{Hermann Edlbauer}
    \affiliation{Universit\'e Grenoble Alpes, CNRS, Grenoble INP, Institut N\'eel, F-38000 Grenoble, France}
\author{Aymeric Richard}
    \affiliation{Universit\'e Grenoble Alpes, CNRS, Grenoble INP, Institut N\'eel, F-38000 Grenoble, France}
\author{Shunsuke Ota}
    \affiliation{Department of Electrical and Electronic Engineering, Tokyo Institute of Technology, Tokyo 152-8550, Japan}
    \affiliation{National Institute of Advanced Industrial Science and Technology (AIST), National Metrology Institute of Japan (NMIJ), 1-1-1 Umezono, Tsukuba, Ibaraki 305-8563, Japan}
\author{Wanki Park}
    \affiliation{Department of Physics, Korea Advanced Institute of Science and Technology, Daejeon 34141, South Korea}
\author{Jeongmin Shim}
    \affiliation{Department of Physics, Korea Advanced Institute of Science and Technology, Daejeon 34141, South Korea}
\author{Arne Ludwig}
    \affiliation{Lehrstuhl f\"{u}r Angewandte Festk\"{o}rperphysik, Ruhr-Universit\"{a}t Bochum, Universit\"{a}tsstra\ss e 150, D-44780 Bochum, Germany}
\author{Andreas D. Wieck}
    \affiliation{Lehrstuhl f\"{u}r Angewandte Festk\"{o}rperphysik, Ruhr-Universit\"{a}t Bochum, Universit\"{a}tsstra\ss e 150, D-44780 Bochum, Germany}
\author{Heung-Sun Sim}
    \affiliation{Department of Physics, Korea Advanced Institute of Science and Technology, Daejeon 34141, South Korea}
\author{Matias Urdampilleta}
    \affiliation{Universit\'e Grenoble Alpes, CNRS, Grenoble INP, Institut N\'eel, F-38000 Grenoble, France}
\author{Tristan Meunier}
    \affiliation{Universit\'e Grenoble Alpes, CNRS, Grenoble INP, Institut N\'eel, F-38000 Grenoble, France}
\author{Tetsuo Kodera}
    \affiliation{Department of Electrical and Electronic Engineering, Tokyo Institute of Technology, Tokyo 152-8550, Japan}
\author{Nobu-Hisa Kaneko}
    \affiliation{National Institute of Advanced Industrial Science and Technology (AIST), National Metrology Institute of Japan (NMIJ), 1-1-1 Umezono, Tsukuba, Ibaraki 305-8563, Japan}
\author{Hermann Sellier}
    \affiliation{Universit\'e Grenoble Alpes, CNRS, Grenoble INP, Institut N\'eel, F-38000 Grenoble, France}
\author{Xavier Waintal}
    \affiliation{Universit\'e Grenoble Alpes, CEA, INAC-Pheliqs, F-38000 Grenoble, France}
\author{Shintaro Takada}
    \affiliation{National Institute of Advanced Industrial Science and Technology (AIST), National Metrology Institute of Japan (NMIJ), 1-1-1 Umezono, Tsukuba, Ibaraki 305-8563, Japan}     
\author{Christopher B\"auerle}
    \altaffiliation{corresponding authors: 	
    \href{mailto: christopher.bauerle@neel.cnrs.fr}{christopher.bauerle@neel.cnrs.fr}}
    \affiliation{Universit\'e Grenoble Alpes, CNRS, Grenoble INP, Institut N\'eel, F-38000 Grenoble, France}

\date{\today}

\begin{abstract}
Electron flying qubits are envisioned as potential information link within a quantum computer \cite{DiVincenzo2000}, but also promise -- alike photonic approaches \cite{OBrien2009} -- a self-standing quantum processing unit \cite{Bauerle2018,Edlbauer2022}.
In contrast to its photonic counterpart, electron-quantum-optics implementations are subject to Coulomb interaction, which provide a direct route to entangle the orbital \cite{Kang2007,Weisz2014} or spin \cite{Barnes2000, Lepage2020, Jadot2021, Krenner2022} degree of freedom.
However, the controlled interaction of flying electrons
at the single particle level has not yet been established experimentally.
Here we report antibunching of a pair of single electrons that is synchronously shuttled through a circuit of coupled quantum rails by means of a surface acoustic wave.
The in-flight partitioning process exhibits a reciprocal gating effect which allows us to ascribe the observed repulsion predominantly to Coulomb interaction.
Our single-shot experiment marks an important milestone on the route to realise a controlled-phase gate for in-flight quantum manipulations.
\end{abstract}

\maketitle

Collision experiments provide fundamental insights into the quantum statistics of elementary particles.
A prime example is the well-known Hong-Ou-Mandel (HOM) interferometer \cite{HOM} where two incident particles are simultaneously scattered at a beam splitter.
For the case of indistinguishable photons, they bunch due to Bose-Einstein statistics leading to an increased probability of the two particles arriving at the same detector.
For colliding electrons, on the other hand, antibunching occurs because of two coexisting mechanisms -- the Pauli exclusion principle and Coulomb repulsion -- causing coincidental counts at the two detectors.
In collision experiments performed within the two-dimensional electron gas (2DEG) in a solid-state device, it is typically assumed that Coulomb interaction is negligible due to screening by the surrounding Fermi sea and, therefore, Pauli exclusion is the dominant repulsion mechanism \cite{Liu1998, Dubois2013, Bocquillon2013}.
Coulomb interaction provides however a direct route for orbital entanglement \cite{Blatter2013}, enabling experiments on quantum nonlocality \cite{Bell1964,Aspect1982} and the implementation of a two-qubit gate for single flying electrons \cite{Bauerle2018,Edlbauer2022,Barnes2000,Ionicioiu2001}.
Whether such a controlled Coulomb interaction is experimentally feasible and sufficient for orbital entanglement, however, has not yet been demonstrated.

In this work, we address this question by implementing the HOM interferometer in a depleted single-electron circuit with coupled quantum rails.
In the absence of the Fermi sea along the transport paths, the screening effect is expected to be significantly reduced.
We move a pair of isolated electrons from two different input ports towards a tunnel-coupled region employing the confinement potential accompanying a surface acoustic wave (SAW) \cite{Hermelin2011, McNeil2011, Delsing2019}.
In order to make the co-propagating electron pair collide, we tune the transmission in this coupling region such that the individual electrons are equally partitioned towards the two outputs.
We synchronise the transport via triggered-sending processes \cite{Takada2019} that we apply independently on each electron source.
This control of the time delay between the two electrons allows us to contrast the full-counting statistics of the single-shot scattering events with and without interaction.
Comparing our experimental results to numerical simulations, we identify the major cause of in-flight interaction and assess its applicability for orbital entanglement.

\begin{figure*}[t]
\includegraphics[width=\dcwidth]{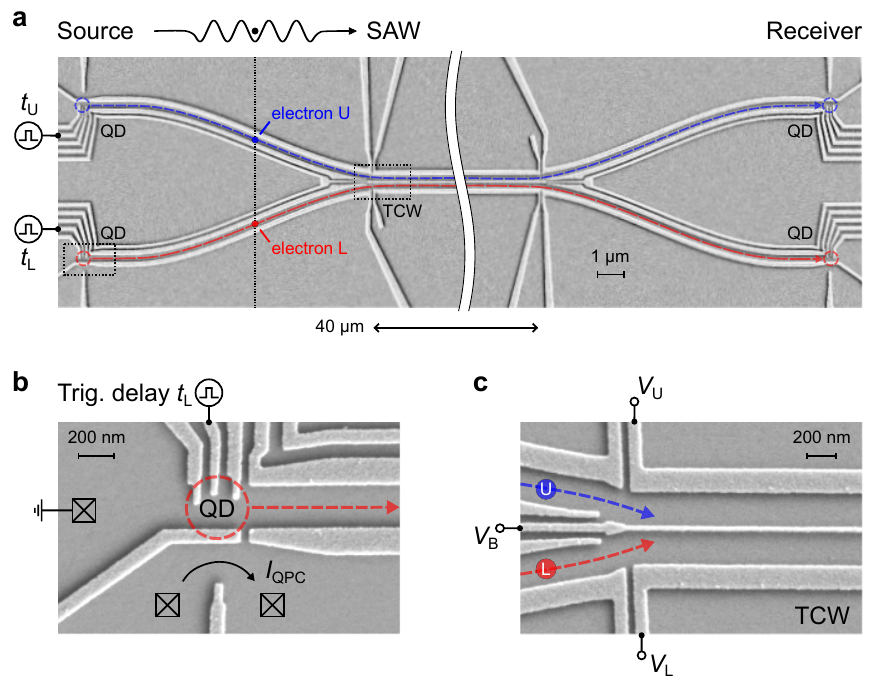}
\caption{Experimental setup.   		
\textsf{\textbf{(a)}}
    Scanning-electron-microscopy (SEM) image of the sound-driven circuit of coupled quantum rails.
    A pair of single electrons (points) is transferred via a SAW train between distant quantum dots (QD) along two quantum rails (dashed lines).
    Along a length of 40~$\mu$m, the two rails form a tunnel-coupled wire (TCW) where they are only separated by a narrow potential barrier.
    \textsf{\textbf{(b)}}
    SEM image of the lower source QD with indication of the electron trajectory (dashed arrow), the electrometer-current (arrow) through the nearby quantum point contact (QPC) and the voltage-pulse trigger of the sending process (with time delay $t_{\rm L}$).
    The crossed boxes indicate ohmic contacts to the two-dimensional electron gas.
    \textsf{\textbf{(c)}}
    SEM image of the TCW entrance with schematic indications of the electron trajectories (dashed lines).
    The potential along this coupling region is controlled via applied voltages on the side gates ($V_{\rm U}$ and $V_{\rm L}$) and the tunnel barrier ($V_{\rm B}$).
    \label{fig:setup}
    }
\end{figure*}

The experimental setup consists of a surface-gate-defined circuit hosting a pair of coupled quantum rails (see Fig.~\ref{fig:setup}a).
The SAW is emitted from a regular interdigital transducer (IDT) that is located around 1.6~mm to the left of the single-electron circuit.
It travels with a speed of 2.86~$\mu$m/ns and has a wavelength of 1~$\mu$m.
When propagating along the nanoscale device, the SAW allows shuttling of a single electron between distant quantum dots (QD) \cite{Hermelin2011,McNeil2011} that are located at the respective ends of the coupled transport paths (see Fig.~\ref{fig:setup}b).
The presence of the electron in a QD is traced via the current flowing through a nearby quantum point contact (QPC) as a non-invasive electrometer.
Enhancing the SAW potential modulation up to a peak-to-peak amplitude of $42\pm13$~meV (see Appendix~\ref{suppl:sawampl}), we ensure that the transported electrons are strongly confined during flight \cite{Edlbauer2021}.
The two injection paths of our single-electron circuit converge to a tunnel-coupled wire (TCW).
Over a length of 40~$\mu$m, the two quantum rails in this region are only separated by a narrow barrier that is defined via a 30-nm-wide surface gate.
Before being projected to the upper (U) or lower (L) output channels, a transported electron experiences thus a flight-time of $\approx 14$~ns in this double-well potential.

\begin{figure*}[t]
\includegraphics[width=\dcwidth]{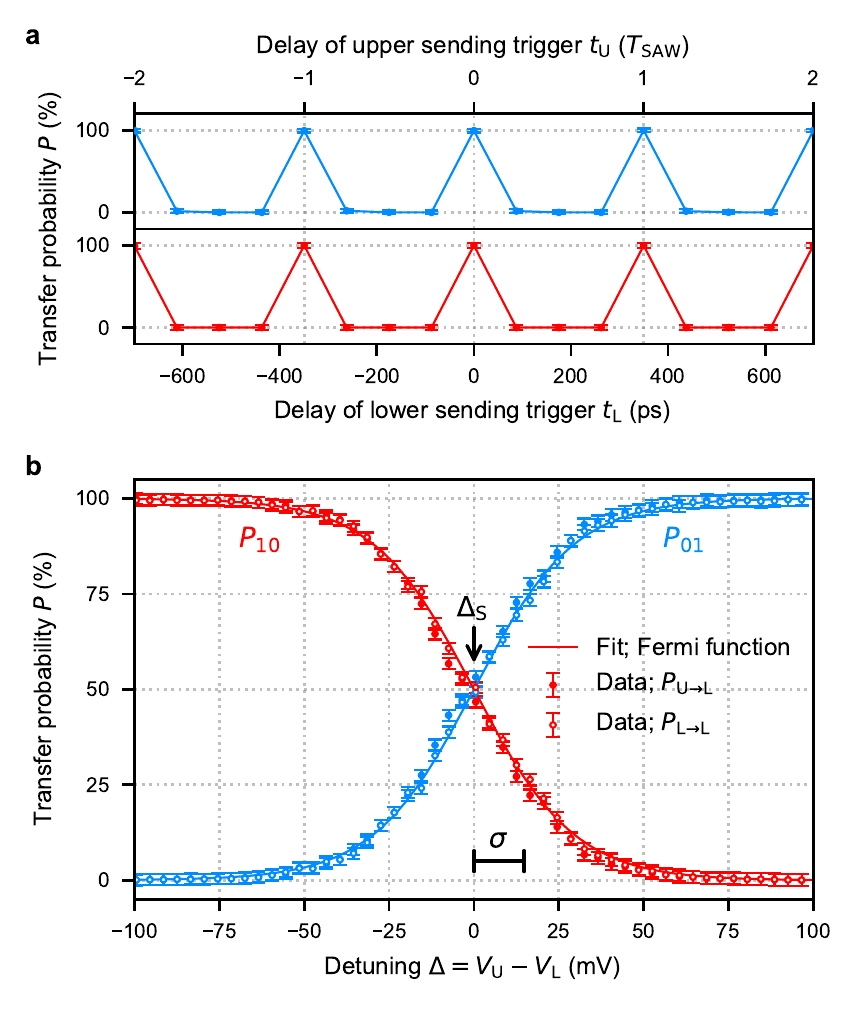}
\caption{Delay-controlled sending and in-flight partitioning.   		
\textsf{\textbf{(a)}}
    Independent measurements of the probability, $P$, of single-electron transport along the upper (lower) quantum rail for different values of the sending-trigger delay $t_{\rm U}$ ($t_{\rm L}$) at the respective source QD.
    The duration of the trigger pulse is $T_{\rm SAW}/4 \approx 90$~ps.
    \textsf{\textbf{(b)}}
    Probability to end up in the upper ($P_{\rm 01}$) or lower ($P_{\rm 10}$) quantum rail for a triggered single-electron emission from the upper (points) and lower (circles) source QD for different potential detuning $\Delta$. 
    The lines show a fit by a Fermi function (see Eq.~\ref{equ:fermi}) with offset $\Delta_{\rm S}$ and width $\sigma$.
    Here, the barrier voltage is set to $V_{\rm B} = -1.10$~V.
    The error bars in both panels are extracted from thousands of single-shot measurements.
    \label{fig:trigdir}
    }
\end{figure*}

A key requirement to realise the HOM interferometer is to control the delay between individual electrons from the two source QDs.
To synchronise the sending process, we apply a 90-ps voltage pulse on the plunger gate of each QD to trigger SAW-driven electron transport on demand \cite{Takada2019}.
In order to characterise the efficiency of this triggering approach, we first tune the voltages on the surface gates into a condition where the two quantum rails are decoupled.
Sweeping the delay of the sending-trigger pulse with respect to the arrival time of the SAW, we observe distinct peaks in the transfer probability as shown in Fig.~\ref{fig:trigdir}a.
The spacing of the peaks coincides with the SAW period $T_{\rm SAW}$ which indicates that we are able to address a specific minimum of the SAW train to transport the electron.
The increase of the transfer probability from $0.35\pm 0.24$\% to $99.77\pm 0.25$\% for both source QDs demonstrates our ability to synchronise the electrons with high accuracy.

To implement the analog of an optical beam splitter for SAW-driven electrons \cite{Ito2021}, we investigate the partitioning of a single flying electron through the coupled quantum rails.
For this purpose, we lower the barrier potential of the TCW such that the electron sent from the upper (lower) source QD can transit into the lower (upper) quantum rail with probability $P_{\rm{U}\to\rm{L}}$ ($P_{\rm{L}\to\rm{U}}$).
To control the in-flight partitioning, we use the side-gate voltages $V_{\rm U}$ and $V_{\rm L}$ to induce a detuning $\Delta = V_{\rm U}-V_{\rm L}$ between the two channels.
Figure \ref{fig:trigdir}b shows transfer probabilities for $V_{\rm B}=-1.10$~V as we detune the double-well potential within the TCW.
We observe a gradual transition which follows a Fermi function:
\begin{equation}
    P_{\rm{i}\to\rm{L}}(\Delta)
    =
    P_{10}(\Delta)
    =
    \frac{1}
    {
    \exp{\big(
    \frac{
    \Delta-\Delta_{\rm{S}}
    }{\sigma}
    \big)}
    + 1
    },
    \label{equ:fermi}
\end{equation}
with $i\in[\rm{U}, \rm{L}]$.
Here, $\Delta_{\rm S}$ indicates the detuning for 50\% transmission -- that is ideally zero for a symmetric device --, and $\sigma$ is the characteristic transition width which is related to the energy distribution of the electron.
Compared to previous work \cite{Takada2019}, we observe a reduced $\sigma$ due to mitigated excitation, which we attribute to the increased SAW confinement \cite{Edlbauer2021} and the improved surface-gate design at the transition region to the TCW employing realistic electrostatic potential simulations \cite{Chatzikyriakou2022}.
To maximise the probability of interaction, it is necessary to prepare an electron pair with similar energy, and thus equal in-flight partitioning in the coupling region.
We find (see Appendix~\ref{suppl:dircoupl}) that such condition is satisfied for $V_{\rm B} \gtrsim -1.15$~V.

\begin{figure*}[t]
\includegraphics[width=\dcwidth]{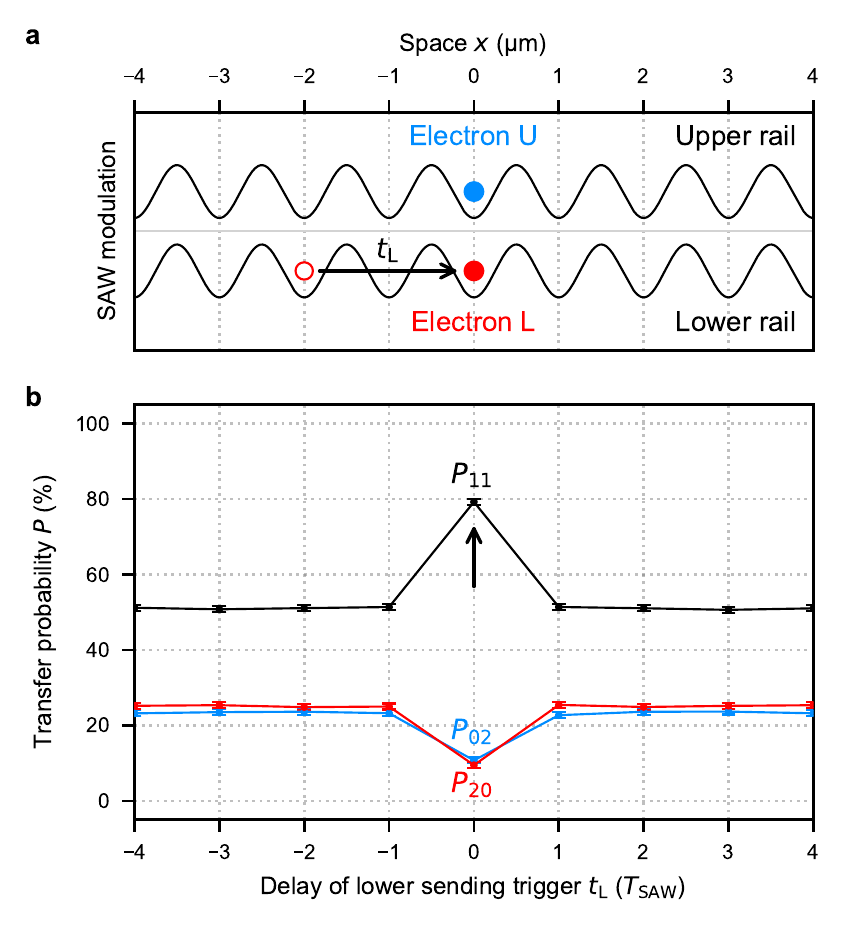}
\caption{Antibunching at synchronised transport.		
    \textsf{\textbf{(a)}}
    Schematic of the collision measurement.
    The delay $t_{\rm U}=0$ of the upper sending trigger (for electron U) is kept fixed while the delay $t_{\rm L}$ of the lower sending trigger (for electron L) is set to successive potential minima of the SAW-train.
    \textsf{\textbf{(b)}}
    Transfer probabilities denoted as $P_{20}$ (both electrons at lower detector), $P_{11}$ (one electron at upper and lower detector) and $P_{02}$ (both electrons at upper detector) as function of the delay $t_{\rm L}$ of electron L.
    Here, the voltage configuration of the TCW is $V_{\rm B} = -1.15$~V and $V_{\rm U} = V_{\rm L} = -1.00$~V.
    The arrow indicates the synchronised condition.
    Each data point is the result from 20.000 single-shot events.
    \label{fig:coldel}
}
\end{figure*}

Before carrying out the collision experiment, we tune the partitioning of each individual flying electron to be 50\% in the coupling region via the voltages $V_{\rm U} = V_{\rm L} = -1.00$~V and $V_{\rm B} = -1.15$~V.
Employing the delays, $t_{\rm U}$ and $t_{\rm L}$, of the sending triggers of the upper and lower source QDs, we control the relative timing between the two transported electrons as sketched in Fig.~\ref{fig:coldel}a.
In particular, we fix the delay of the upper electron ($t_{\rm U} = 0$) and step the delay for the lower triggering pulse in multiples of the SAW period ($t_{\rm L} = k\cdot T_{\rm SAW}$  where $k \in	\mathbb{Z}$) in order to address different SAW minima for transport.
If the electrons tunnel without experiencing the presence of the other, the probabilities at the detectors would follow a Poissonian distribution. 
In this case, we expect 50\% for the probability $P_{11} \equiv P_{\rm{UL}\to\rm{UL}}$ to find one electron in both the upper and the lower detector, and, accordingly, $P_{20} \equiv P_{\rm{UL}\to\rm{LL}}$ and $P_{02} \equiv P_{\rm{UL}\to\rm{UU}}$ to be 25\%.
Figure \ref{fig:coldel}b shows such a measurement of the antibunching probability $P_{11}$ as a function of the trigger delay $t_{\rm L}$ of the electron sent from the lower source QD.
We find $P_{11}\approx 50$\% as expected when the two electrons are transported in different SAW minima ($t_{\rm L} \neq 0$).
As the sending triggers are synchronised ($t_{\rm L} = 0$) and the electron pair is thus sent within the same SAW minimum, we observe in contrast a significant increase of $P_{11}$ up to 80\% resulting from the interaction between the two electrons.
The distinct $P_{11}$ peak underpins our expectation that the flying electrons remain within the initially addressed SAW minimum during transport.
Our observation further indicates that beyond a distance of one SAW period ($\approx 1$~$\mu$m) the interaction of the electron pair gets negligible.

\begin{figure*}[t]
\includegraphics[width=\dcwidth]{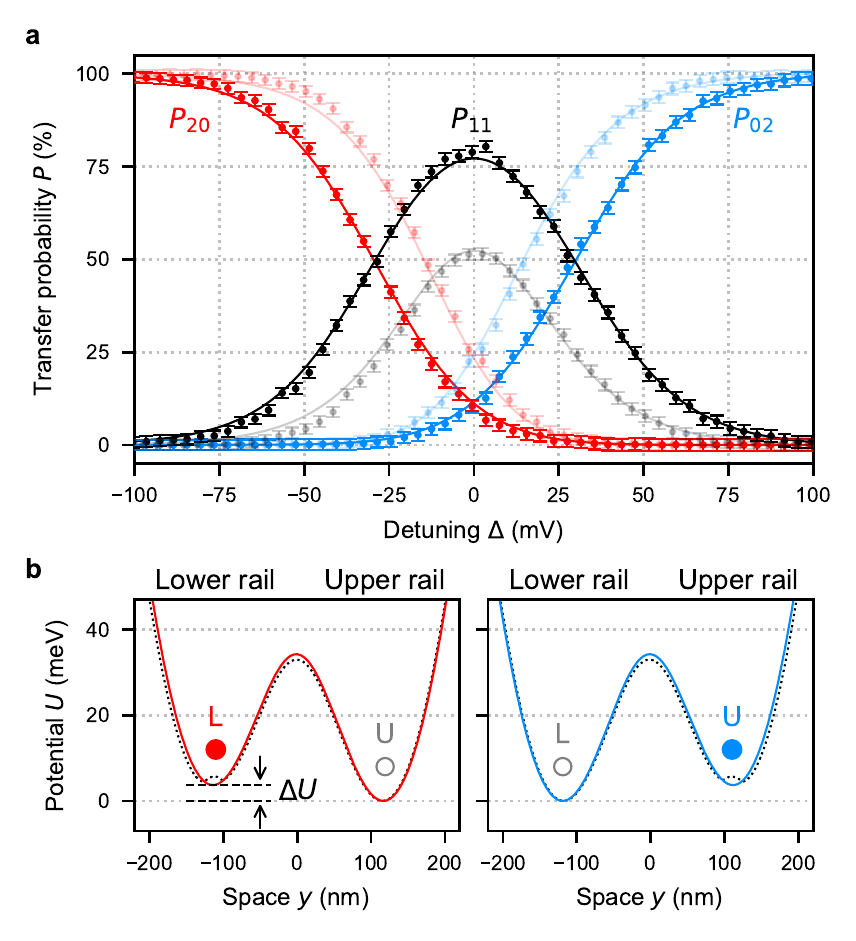}
\caption{Coulomb induced detuning and electron-pair partitioning.	
    \textsf{\textbf{(a)}}
    Measurements of single-shot probabilities for transfer of the electron pair in the same (solid points; $t_{\rm U}-t_{\rm L}=0$) and different (semi-transparent points; $t_{\rm U}-t_{\rm L}=5\cdot T_{\rm SAW}$) potential minima accompanying the SAW train.
    The lines show the results using the Bayesian model -- see Appendix~\ref{suppl:bayes} -- with (solid) and without (semi-transparent) the effective voltage detuning $\delta$ extracted in (b), corresponding respectively to the interacting and non-interacting case.
    \textsf{\textbf{(b)}}
    Detuned potential landscapes observed by one electron (filled point) due to the presence of another (circle).
    The dashed line in the left (right) pannel shows the potential from electrostatic simulations of the symmetrically polarised TCW ($V_{\rm B} = -1.15$~V and $V_{\rm U} = V_{\rm L} = -1.00$~V) with an electron inserted in the lower (upper) coupled transport channel.
    The solid line shows an equivalent potential profile formed by employing a voltage detuning of $\delta = V_{\rm U}-V_{\rm L}\approx \pm 18.5$~mV.
    This electron-gating effect lifts the corresponding side of the double-well potential by $\Delta U \approx 3.7$~meV.
    \label{fig:coldet}
}
\end{figure*}

In order to investigate the nature of the antibunching effect -- Pauli exclusion or Coulomb repulsion --, we perform the partitioning experiment by varying the detuning of the TCW, from the previous symmetric case to the fully detuned situation with the electron pair forced in the same channel.
As reference, we first consider the non-interacting case shown in Fig.~\ref{fig:coldet}a by the semi-transparent data obtained with the two electrons travelling in different SAW minima ($\Delta t = 5\cdot T_{\rm SAW}$).
The observed probabilities are a direct consequence of the partitioning distribution of the individual electrons shown in Fig.~\ref{fig:trigdir}b.
Since the electrons do not interact, the probability to find both electrons in the lower channel is simply the product of the single-electron cases, $P_{20} = P_{\rm{U}\to\rm{L}} \cdot P_{\rm{L}\to\rm{L}}$.
Similarly, we have $P_{02} = P_{\rm{U}\to\rm{U}} \cdot P_{\rm{L}\to\rm{U}}$, and $P_{11} = P_{\rm{U}\to\rm{L}} \cdot P_{\rm{L}\to\rm{U}} + P_{\rm{U}\to\rm{U}} \cdot P_{\rm{L}\to\rm{L}} = 1 - P_{20} - P_{02}$ due to charge conservation. 
The semi-transparent lines indicate the course resulting from this non-interacting model that shows good agreement with the experimental data.
As we send the two electrons synchronously within the same SAW minimum (non-transparent data), we observe a change in the functional course of $P_{20}$ and  $P_{02}$ leading to a significant increase and broadening of $P_{11}$ compared to the non-interacting case.


To find out the physical effect that causes the observed in-flight partitioning of the two interacting electrons, we focus on the Coulomb potential that is experienced by one electron due to the presence of the other.
We perform three-dimensional electrostatic simulations \cite{Chatzikyriakou2022} taking into account the geometry and electronic properties of the presently investigated nanoscale device -- see Methods.
For the sake of simplicity, we consider a symmetric configuration of the surface gate voltages ($V_{\rm U}=V_{\rm L} = -1.00$~V and $V_{\rm B} = -1.15$~V).
Figure \ref{fig:coldet}b shows the result of an electrostatic simulation (dotted line) by adding the density of an electron-charge in the lower or upper rail.
We observe that the double-well potential is tilted by the presence of the electron with an induced asymmetry of 3.7~meV, which can be reproduced by considering an effective gate-voltage detuning $\delta\approx 18.5\pm0.4$~mV (solid line).
Therefore, these numerical results indicate that the electron in the lower rail (L) experiences a potential landscape that is effectively detuned due to the presence of the electron in the upper rail (U), and {\it vice versa}.

To model the two-electron partitioning process with interaction, we include such a reciprocal electron-gating effect (parameterized by $\delta$) in the single-electron partitioning distribution (see Eq.~\ref{equ:fermi}) as $P_{i\to\rm{j}}(\Delta \pm \delta)$ where $i,j\in[\rm{U},\rm{L}]$.  
In combination with the Bayes' theorem, we derive -- see Appendix~\ref{suppl:bayes} -- the following expression:
\begin{equation}
    P_{20}(\Delta) = 
    \frac{
    P_{\rm{L}\to\rm{L}}(\Delta+\delta) \cdot P_{\rm{L}\to\rm{L}}(\Delta-\delta)
    }{
    \frac{P_{\rm{L}\to\rm{L}}(\Delta+\delta)}{P_{\rm{U}\to\rm{L}}(\Delta+\delta)}
    + P_{\rm{L}\to\rm{L}}(\Delta-\delta)
    - P_{\rm{L}\to\rm{L}}(\Delta+\delta)
    },
    \label{equ:p20int}
\end{equation}
which allows us to construct $P_{02}(\Delta)$ and $P_{11}(\Delta)$.
The solid lines shown in Fig.~\ref{fig:coldet}a indicate the courses of $P_{20}$, $P_{11}$ and $P_{02}$ resulting from Eq. \ref{equ:p20int} with $\delta=18.5$~mV, and $P_{\rm{L}\to\rm{L}}$ and $P_{\rm{U}\to\rm{L}}$ extracted from the individual non-interacting partitioning data.
Since the Bayesian model is solely based on electrostatics, the excellent agreement with the experimental data without adjustable parameters indicates that the Coulomb interaction is the major source of the increased antibunching probability.
We further verify this conclusion by performing exact diagonalization calculations -- see Appendix~\ref{suppl:ed} -- in which the long-range Coulomb repulsion is taken into account, and find a good quantitative agreement both on the increased antibunching probability and on the increased transition width.

\begin{figure*}[t]
\includegraphics[width=\dcwidth]{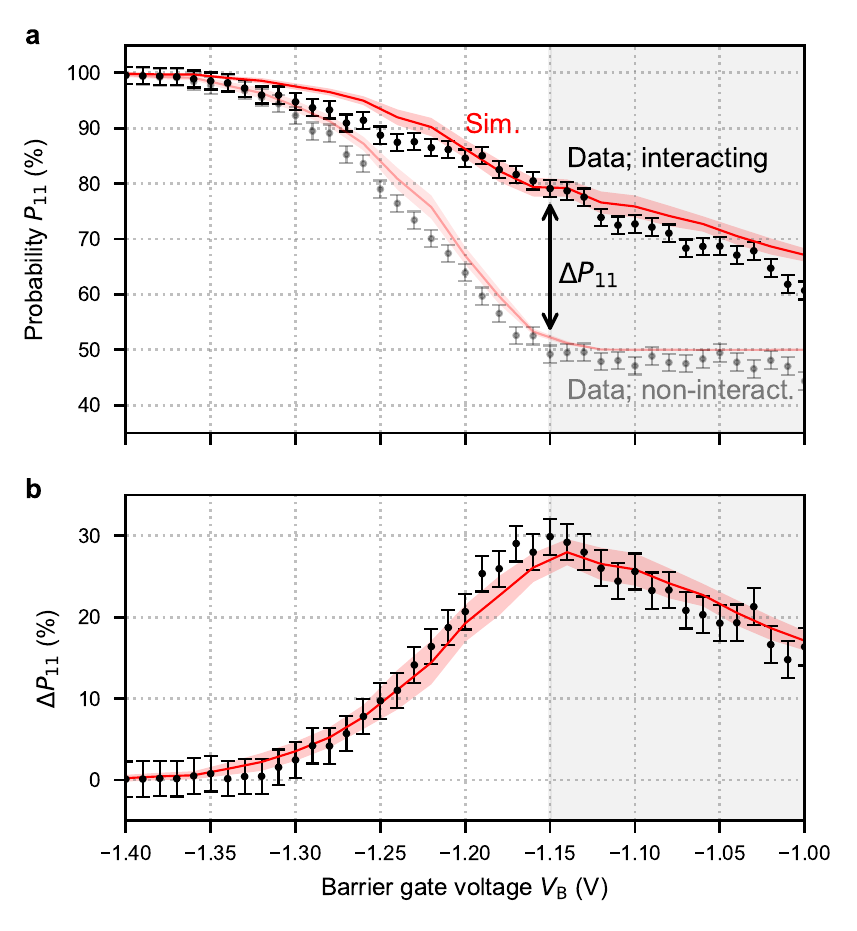}
\caption{Barrier dependence of antibunching rate.
    \textsf{\textbf{(a)}}
    Antibunching probability $P_{11}$ for transport in the same (black points; interacting) and different (grey points; non-interacting) SAW minima as function of the voltage $V_{\rm B}$ applied on the barrier gate.
    The line shows the course of the Bayesian model with an associated error (semi-transparent area) which stems from the deduction of the Coulomb-equivalent detuning $\delta$ from experimental data.
    The shaded region ($V_{\rm B}\geq-1.15$~V) highlights the regime where $P_{11}$ of the non-interacting case is saturated at $\approx50$\%.
    \textsf{\textbf{(b)}}
    Excess of antibunching rate $\Delta P_{11}$ resulting from the electron-pair interaction.
    \label{fig:tcw}
}
\end{figure*}

Having identified Coulomb interaction as the main cause of antibunching for a specific configuration, we now check whether this assertion also holds when the barrier potential is changed.
For this purpose, we investigate the antibunching probability $P_{11}$ at a symmetric detuning ($\Delta=0$) as a function of the barrier gate voltage $V_{\rm B}$ (see Fig.~\ref{fig:tcw}a).
Focusing on the non-interacting case (semi-transparent data), increasing the barrier height ($V_{\rm B}<-1.15$~V) reduces the transmission of each electron to the opposite channel, leading to a gradual increase of $P_{11}$ above 50\% and up to 100\% when both rails are fully separated.
This regime of barrier voltages with progressively decoupled rails is therefore not suitable to investigate the influence of the electron-pair interaction solely.
When the electron pair is transported synchronously (black data), we observe a similar increase of $P_{11}$ in this regime starting from the optimal value of 80\% discussed previously.
For lower barrier heights ($V_{\rm B}>-1.15$~V; grey area), the antibunching probability $P_{11}$ decreases gradually below 80\% while the non-interacting data is saturated at 50\%.
To model this dependence on the barrier height, we extract the Coulomb-equivalent detuning $\delta$ from two-electron partitioning experiments performed at three different barrier voltages $V_{\rm B}\in\{-1.150,-1.125,-1.100\}$~V (see Appendix~\ref{suppl:delta}).
Using a linear course of $\delta(V_{\rm B})$, the simulation from the Bayesian model (red) shows excellent agreement with the experimental data.
The quantitative comparison indicates that Coulomb interaction is dominant for a wide range of barrier voltages.

Next, we address the question of what limits the maximum observed antibunching probability at $P_{11}\approx80$\%.
A possible explanation could be the occupation of excited states by the flying electrons \cite{Takada2019}.
If their energy overcomes the Coulomb repulsion, $P_{11}$ is expected to be reduced.
To check this possibility, we numerically investigate the effect of excitation in the antibunching process using the Bayesian model -- see Appendix~\ref{suppl:p11}.
We find that $P_{11}$ is expected to exceed 99\% if the effective thermal excitation of the electron is reduced from the present 3~meV to below 1~meV.


For the implementation of the two-qubit gate with flying electrons \cite{Barnes2000,Lepage2020, Helgers2022},
let us estimate the extent of the reciprocal phase shift, $\varphi=U_{\rm C} \cdot t/\hbar$, induced on the wavefunctions of the electron pair after an interaction time $t$.
The energy due to the Coulomb interaction is represented here as $U_{\rm C}(r)=\frac{e^2}{4\pi\varepsilon_0\varepsilon_r} \frac{1}{r}$ where $r$ is the distance between the two electrons, $\varepsilon_0$ is the vacuum permittivity and $\varepsilon_r=12.88$ is the dielectric constant of GaAs.
From potential simulations, we extract a distance of $r\approx 230$~nm, which gives a Coulomb energy $U_{\rm C} \approx 0.5$~meV.
Considering the SAW velocity $v_{\rm SAW} \approx 2.86$~$\mu$m/ns, we expect a phase rotation $\varphi=\pi$ (Bell state formation) over a propagation distance $l=\pi\hbar v_{\rm SAW}/U_{\rm C}\approx 12$~nm.
This estimation shows that in-flight Coulomb interaction within a TCW introduces a significant reciprocal phase shift capable of entangling the orbits in a SAW-driven single-electron circuit.


In conclusion, we have demonstrated the controlled interaction between two single flying electrons transported by sound.
This has been achieved through the implementation of the HOM interferometer with a circuit of coupled quantum rails.
Synchronising the transport of a pair of individual electrons, we witnessed single-shot events of fermionic antibunching.
To address the underlying mechanism, we performed quantitative electrostatic simulations, and observed a reciprocal electron-gating effect.
Developing a Bayesian model, which contains no adjustable parameter, we showed quantitative agreement with the entire set of two-electron collision data.
This provides strong evidence that the observed antibunching is mediated by Coulomb repulsion.
Further estimating the strength of this Coulomb interaction, we highlight that it is more than sufficient for the formation of a fully entangled Bell state.
Combining this controlled interaction with novel, scalable single-electron-transport techniques \cite{Wang2022}, our results set an important milestone towards the implementation of the controlled-phase gate for SAW-driven flying electron qubits.

\section*{Methods}
\label{sec:methods}


{\bf SAW transducer.}
The employed IDT consists of 111 cells of period $\lambda_0=1$~$\mu$m. 
The resonance frequency is $f_0=v_{\rm SAW}/\lambda_0\approx 2.86$~GHz at cryogenic temperatures.
To reduce internal reflections at resonance, we employ a double-electrode pattern for the transducers. 
The surface electrodes of the IDTs are fabricated using standard electron-beam lithography with successive thin-film evaporation (Ti 3~nm, Al 27~nm) on GaAs/AlGaAs heterostructure.
The transducer has an aperture of 30~$\mu$m with the SAW propagation direction along $[1\bar{1}0]$.
For single-electron transport, we employ an input signal at the resonance frequency with a duration of 50~ns.
To achieve strong SAW confinement, the input signal for SAW formation is enhanced by a high-power amplifier (ZHL-4W-422+; +25~dB) prior injection.

{\bf Electron-transport experiments.} 
We use a Si-modulation-doped GaAs/AlGaAs heterostructure grown by molecular beam epitaxy (MBE). The two-dimensional electron gas (2DEG) is located 110~nm below the surface, with an electron density of $n\approx 2.8 \times 10^{11}$~$\rm cm^{-2}$ and a mobility of $\mu\approx 9\times 10^5$~$\rm cm^{2}V^{-1}s^{-1}$. 
Metallic surface gates (Ti 3~nm, Au 14~nm) define the nanostructures. 
The experiment is performed at a temperature of about 20~mK in a $^3\textrm{He}/^4\textrm{He}$ dilution refrigerator. 
At low temperatures, the 2DEG below the transport channels and the QDs are completely depleted via a set of negative voltages applied on the surface gates. 
To enable triggering of the sending process, the plunger gate of each source QD is connected to a broadband bias tee (SHF AG; 20~kHz to 40~GHz).

{\bf Synchronisation between SAW emission and triggered sending process.}
We employ two dual-channel arbitrary waveform generators (AWG, Keysight M8195A) synchronised via an synchronisation unit (Keysight M8197A) for the antibunching experiments. 
A small jitter of $\approx 1$~ps between the AWG channels allows to control precisely the timing between SAW emission and the triggered sending process at each QD.

{\bf Potential simulations.}
The simulations are performed with the commercial Poisson solver nextnano \cite{Birner2007}.
We define a three-dimensional structure with realistic heterostructure layers and gate geometries where the corresponding materials' properties are taken into account.
In our electrostatic model, a metallic gate is expressed as a Schottky barrier \cite{Chatzikyriakou2022,Hou2018}.
On the free surface, a layer of surface charges simulates the Fermi-level-pinning effect that is well-known in GaAs substrates \cite{Sze2006}. 
Using one-dimensional simulations, we first calibrate the dopant concentration with a surface gate such that it reproduces the 2DEG density at the interface of GaAs and AlGaAs.
We then adjust similarly the surface charges in the absence of the gate.
The presence of an electron in one side of the rail is emulated by inserting the charge of an electron in a volume of $\Delta x=150$~nm, $\Delta y=17$~nm and $\Delta z=1$~nm, where $x$ ($y$) is parallel (perpendicular) to the SAW propagation direction, and $z$ is the growth direction of the heterostructure.

\begin{acknowledgments}
We acknowledge fruitful discussions with Vyacheslavs Kashcheyevs and Elina Pavlovska.
J.W. acknowledges the European Union's Horizon 2020 research and innovation program under the Marie Skłodowska-Curie grant agreement No 754303.
A.R. acknowledges financial support from ANR-21-CMAQ-0003, France 2030, project QuantForm-UGA.
T.K. and S.T. acknowledge financial support from JSPS KAKENHI Grant Number 20H02559.
W.P., J.S., and H.-S.S. acknowledge support from Korea NRF via the SRC Center for Quantum Coherence in Condensed Matter (Grant No. 2016R1A5A1008184).
C.B.  acknowledges financial support from the French Agence Nationale de la Recherche (ANR), 
project QUABS ANR-21-CE47-0013-01.
This project has received funding from the European Union’s H2020 research and innovation program under grant agreement No 862683 "UltraFastNano".
\end{acknowledgments}

\appendix

\section{Estimation of SAW amplitude}
\label{suppl:sawampl}

In order to investigate how the input RF power $P_{\rm in}$ applied on the IDT relates to the SAW peak-to-peak amplitude $A_{\rm SAW}$, we measure the SAW-induced modulation of Coulomb-blockade resonances of a QD.
For this purpose, we polarise the surface gates such that the QD is not depleted, and apply a bias voltage $V_{\rm SD}$ across the two leads.
Varying $V_{\rm SD}$ as a function of the plunger gate voltage $V_{\rm P}$, we measure the conductance across the QD and obtain Coulomb diamonds as shown in Fig.~\ref{sfig:sawampl}a.
This data allows us to extract the quantum dot's charging energy $E_{\rm C}$ and the voltage spacing $V_{\rm C}$ between Coulomb-blockade peaks.
The voltage-to-energy conversion factor is thus $\eta = E_{\rm C}/V_{\rm C} \approx 0.05 \pm 0.01$~eV/V.
Knowing $\eta$, we can now deduce the SAW amplitude $A_{\rm SAW}$ from a given input power $P_{\rm in}$ via the relation:
\begin{equation}
A_{\rm SAW} \;\text{[eV]} = 2 \cdot \eta \cdot 10^{(P_{\rm in}\;\text{[dBm]}-P_0)/20}\text{,}
\label{equ:ampl}
\end{equation}
where $P_0$ is a fit parameter accounting for power losses.
$P_0$ is determined by comparison of Eq.~\ref{equ:ampl} to the SAW-induced broadening of the Coulomb-blockade resonances.
Figure \ref{sfig:sawampl}b shows a conductance measurement as function of $V_{\rm P}$ and $P_{\rm in}$ for $V_{\rm SD}\approx 20$~$\mu$V.
The data shows Coulomb-blockade peaks that broaden according to Eq.~\ref{equ:ampl} with $P_0\approx 36.8 \pm 0.3$~dBm as indicated by the solid lines (the dashed lines represent the error margin).
Considering the here-employed input power of $P_{\rm in}\approx 28$~dBm, we extrapolate an amplitude of $A_{\rm SAW}\approx 42 \pm 13$~meV.
This value lies beyond the amplitude $A_{\rm SAW}\approx 17 \pm 8$~meV reported from our previous device of coupled quantum rails \cite{Takada2019} (lower horizontal line) which indicates that we have successfully improved the SAW confinement.
To estimate if the transported electron would stay within a SAW minimum, we compare the present $A_{\rm SAW}$ with the 95\% confinement threshold of $A_{\rm SAW}\approx 24$~meV that was deduced from time-of-flight measurements along a straight quantum rail \cite{Edlbauer2021} (upper horizontal line).
Our results indicate that the employed SAW power is strong enough to ensure in-flight confinement.

\section{Barrier dependence of single-electron partitioning}
\label{suppl:dircoupl}

The partitioning data from a single electron follows a Fermi function (see Eq.~\ref{equ:fermi}) with a half-transmission detuning $\Delta_{\rm S}$ and a characteristic transition width $\sigma$.
Figure \ref{sfig:dircoup}a shows the evolution of $\Delta_{\rm S}$ as a function of the barrier-gate voltage $V_{\rm B}$ for an electron sent from the upper (solid line) or lower (dashed line) source QD.
For a large barrier height ($V_{\rm B} < -1.15$~V), an asymmetric polarisation of the channel gates ($\Delta_{\rm S} \neq 0$) is required to achieve 50\% transmission.
Comparing in-flight partitioning data from an individual electron injected from each source QD, we observe that $\Delta_{\rm S}$ converges gradually to a matching value when $V_{\rm B}$ becomes more positive (lower barrier).
The course of the transition width $\sigma$ for both injection sides shows identical behaviour (see Fig.~\ref{sfig:dircoup}b).
Since $\sigma$ is related to the energy state of the partitioned electron, we find a minimal excitation for $V_{\rm B}$ between $\approx -1.15$~V and -1.05~V.

\section{Bayesian model of in-flight partitioning mediated by Coulomb interaction}
\label{suppl:bayes}

In the following, we employ Bayesian probability calculus to derive the two-electron collision probabilities $P_{20}$, $P_{11}$ and $P_{02}$ from the single-electron-partitioning data $P_{i \to j}$ for $i,j \in [\rm{U}, \rm{L}]$.
Here, a transported electron enters from the input $i$ and exits at the output $j$.
In the case of two transported electrons, the probability to find both electrons in L can be defined via the joined probability
\begin{equation}
    P_{20} \equiv P_{\rm{UL}\to \rm{LL}} = P_{\rm{L}\to\rm{L}|\rm{U}\to\rm{L}} \cdot P_{\rm{U}\to\rm{L}}
    \label{equ:joined_p20}
\end{equation}
where $P_{\rm{L}\to\rm{L}|\rm{U}\to\rm{L}}$ is the conditional probability to find the electron sent from L at the exit L when electron U is present in channel L, and $P_{\rm{U}\to\rm{L}}$ is the probability to find electron U in channel L independent on the location of electron L.

Expressing $P_{\rm{L}\to\rm{L}|\rm{U}\to\rm{L}}$ and $P_{\rm{L}\to\rm{L}|\rm{U}\to\rm{U}}$ via the Bayes' theorem $P_{A|B} = \frac{P_{B|A}\cdot P_{A}}{P_{B}}$, and knowing $P_{\rm{U}\to\rm{U}} = 1 - P_{\rm{U}\to\rm{L}}$ due to charge conservation, we derive $P_{\rm{U}\to\rm{L}}$ as:
\begin{equation}
    P_{\rm{U}\to\rm{L}} = 
    \frac{P_{\rm{L}\to\rm{L}|\rm{U}\to\rm{U}} \cdot P_{\rm{U}\to\rm{L}|\rm{L}\to\rm{L}}}
    {P_{\rm{L}\to\rm{L}|\rm{U}\to\rm{U}} \cdot P_{\rm{U}\to\rm{L}|\rm{L}\to\rm{L}}+
    P_{\rm{U}\to\rm{U}|\rm{L}\to\rm{L}} \cdot P_{\rm{L}\to\rm{L}|\rm{U}\to\rm{L}}}.
    \label{equ:pbetal}
\end{equation}
Note that here $P_{\rm{U}\to\rm{L}}$ does not need to be equivalent to the single-electron case due to the mutual influence between the electrons. 

Let us first focus on the non-interacting case where the two electrons do not influence each other. 
For two independent events, the conditional probability satisfies $P_{A|B}=P_A$.
Applying this relation to equations \ref{equ:joined_p20} and \ref{equ:pbetal}, we obtain
\begin{equation}
    P_{20} = P_{\rm{L}\to\rm{L}} \cdot P_{\rm{U}\to\rm{L}}
\end{equation}
that follows the Poisson binomial distribution.

In the interacting case, the presence of electron U influences L, and \textit{vice versa}. 
For the presently studied experimental configuration, our potential simulations indicate that the Coulomb potential of electron U effectively detunes the potential landscape that is observed by L.
The effect is equivalent to an effective voltage detuning on the surface gates by $\delta$.
Including this Coulomb interaction, we find:
\begin{align}
    P_{\rm{L}\to\rm{L}|\rm{U}\to\rm{U}} &= 1-P_{\rm{L}\to\rm{U}|\rm{U}\to\rm{U}} = P_{\rm{L}\to\rm{L}}(\Delta -\delta) \\
    P_{\rm{L}\to\rm{L}|\rm{U}\to\rm{L}} &= 1-P_{\rm{L}\to\rm{U}|\rm{U}\to\rm{L}} = P_{\rm{L}\to\rm{L}}(\Delta +\delta).
\end{align}
Similarly, the influence of electron L on electron U is expressed as
\begin{align}
    P_{\rm{U}\to\rm{L}|\rm{L}\to\rm{U}} &= 1-P_{\rm{U}\to\rm{U}|\rm{L}\to\rm{U}} = P_{\rm{U}\to\rm{L}}(\Delta -\delta) \\
    P_{\rm{U}\to\rm{L}|\rm{L}\to\rm{L}} &= 1-P_{\rm{U}\to\rm{U}|\rm{L}\to\rm{L}} = P_{\rm{U}\to\rm{L}}(\Delta +\delta).
\end{align}

Substituting these relations in Eq.~\ref{equ:joined_p20} and \ref{equ:pbetal}, we obtain the joined probability
\begin{equation}
    P_{20}(\Delta) = 
    \frac{
    P_{\rm{L}\to\rm{L}}(\Delta+\delta) \cdot P_{\rm{L}\to\rm{L}}(\Delta-\delta)
    }{
    \frac{P_{\rm{L}\to\rm{L}}(\Delta+\delta)}{P_{\rm{U}\to\rm{L}}(\Delta+\delta)}
    + P_{\rm{L}\to\rm{L}}(\Delta-\delta)
    - P_{\rm{L}\to\rm{L}}(\Delta+\delta)
    }
\end{equation}

Following the same procedure, we can construct $P_{02}(\Delta)$ and $P_{11}(\Delta)$ from
\begin{align}
	P_{02} &\equiv P_{\rm{UL}\to \rm{UU}} = P_{\rm{L}\to\rm{U}|\rm{U}\to\rm{U}} \cdot P_{\rm{U}\to\rm{U}} 
	\label{equ:joined_p02} \\
	P_{11} &\equiv P_{\rm{UL}\to \rm{UL}} = 1 - P_{20} - P_{02}.
    \label{equ:joined_p11}
\end{align}

\section{Exact diagonalization for transfer probabilities}
\label{suppl:ed}

Here we calculate the two-electron transfer probabilities based on the exact diagonalization method, in which the potential shape of the moving QDs induced by the SAW train, the Coulomb interaction between electrons, and an ensemble described by an effective temperature are taken into account.
The results are in qualitatively good agreement with the experimental data of the transfer probabilities as a function of the detuning $\Delta$, the input power $P_{\rm{in}}$ for SAW generation, and the barrier gate voltage $V_{\rm B}$.
This supports that the experimental findings of the antibunching behavior originate from Coulomb interaction.

In this model, we consider a system with two electrons confined in a two dimensional potential 
$U(x,y) = U_{\textrm{SAW}}(x) + U_{\textrm{QDs}}(y)$ where $x$ ($y$) corresponds to the parallel (perpendicular) direction with respect to the transport channel. 
$U_{\textrm{SAW}}(x)$ is the confinement potential along the SAW propagation direction $x$ which we define as
\begin{equation}
	U_{\text{SAW}}(x) = \dfrac{A_{\rm SAW}}{2} \left(1- \cos\left({2\pi\frac{x}{\lambda}}\right)\right),
	\label{equ:USAW}
\end{equation}
where $\lambda \approx 1\textrm{ $\mu$m}$ is the SAW period, and $A_{\rm SAW}$ is the peak-to-peak SAW amplitude determined by the input power $P_{\rm in}$ (see Eq.~\ref{equ:ampl}). 
We derive accordingly the SAW confinement energy $\hbar\omega_x$ using the Taylor expansion of Eq.~\ref{equ:USAW} at the local minimum $x=0$ as
\begin{equation}
\omega_x(P_{\rm in}) = \left ( \frac{\pi}{\lambda} \right ) \left ( \frac{2}{m_{e}} \right )^{\frac{1}{2}}  \left ( A_0 10^{\frac{P_{\rm in}}{20}} \right ) ^{\frac{1}{2}},
\label{equ:omega_x_P}
\end{equation}
where $m_e \approx 0.067 m_0$ is the effective electron mass in GaAs and $A_0\approx1.7$~meV is the peak-to-peak amplitude at $P_{\rm in}=0$~dBm.
For the experimental condition of $P_{\rm in}=28$~dBm, we estimate $\hbar\omega_x \approx 1$~meV.

On the other hand, $U_{\textrm{QDs}}(y)$ describes the double-QD potential generated by a single minimum of the SAW train and the barrier gate voltage $V_{\rm B}$.
This potential along the transverse direction $y$ is modelled, following Ref.~\cite{Burkard1999}, as
\begin{equation}
	U_{\textrm{QDs}}(y) = \frac{1}{2} m_e \omega_y^2 \frac{1}{(2d)^2} (y^2 - d^2)^2 - \frac{\alpha \Delta}{2d} y,
\end{equation}
where $\omega_y$ corresponds to the single-particle level spacing in the transverse direction $y$, $2d$ is the distance between the two QDs, and the detuning $\Delta$ combined with the conversion factor $\alpha$ determines the asymmetry of this double-well potential.
From electrostatic simulations (see Methods) using experimental conditions ($V_{\rm B} = -1.15$~V and $V_{\rm U} = V_{\rm L} = -1.00$~V), we extract $\hbar \omega_y = 3.6$~meV and $2d = 230$~nm.

To include the interaction between the electron pair, we consider the Coulomb energy
\begin{equation}
	U_{\rm C}(r) = \frac{e^2}{4 \pi \varepsilon_0 \varepsilon_r } \frac{1}{(r^2 + d_{\rm{min}}^2)^{1/2}}.
\end{equation}
Here, $r$ is the distance between the electrons, $\varepsilon_r = 12.88$ is the dielectric constant of GaAs, $\varepsilon_0$ is the vacuum permittivity, $e$ is the electron charge, and $d_{\rm{min}}$ is the width of the quantum well that confines the two-dimensional electron gas in $z$ direction. Considering the quantum-well confinement width, we choose $d_{\rm{min}} = 5$~nm.

Next, we define the Hamiltonian for such a system of two interacting electrons as $H = (p_{x,1}^2 + p_{y,1}^2)/ (2m_e) + (p_{x,2}^2 + p_{y,2}^2)/(2m_e) + U(x_1,y_1) + U(x_2,y_2)+ U_{\rm C}(r)$, where $(p_{x,i}, p_{y,i})$ are the momentum in $x$ and $y$ direction, respectively, of electron $i=1,2$ and $(x_i,y_i)$.
To solve the Hamiltonian, we convert the two dimensional continuous space into a rectangular discrete lattice, and apply the exact diagonalization method. 
We use a grid of $20 \times 32$ with lattice constants $\delta x \approx 26$~nm and $\delta y \approx 13$~nm. 
Note that $\delta x$ and $\delta y$ are shorter than the characteristic length scale for the QDs, $l_x = \sqrt{\hbar / (m_e \omega_x)}$ and $l_y = \sqrt{\hbar / (m_e \omega_y)}$, respectively.
Here, $\omega_x$ is determined by Eq.~\ref{equ:USAW}. 
To reduce computational costs in calculating two-electron eigenstates, we discard single-particle basis states whose energy is higher than $8 k_B T_{\rm{eff}}$, where $k_B$ is the Boltzmann constant and $T_{\rm{eff}}$ is the effective temperature discussed later.
Under these conditions, the calculated probabilities are converged within 1\% error when decreasing $\delta x$, $\delta y$ or the number of truncated states.

To better describe the experimental condition, we include an effective thermal temperature $T_{\rm{eff}}$ in the system to represent the nonequilibrium state due to nonadiabatic excitations generated during the electron transport \cite{Takada2019}.
Note that $T_{\rm{eff}}$ is different from the electron temperature of the experiment.
For this purpose, we construct a thermal ensemble by using the eigenstates of the Hamiltonian $H$ obtained by the exact diagonalization and the thermal Boltzmann factor.
The density operator of the ensemble is written as $\hat{\rho} = \hat{\rho}_{\rm s}/4 + 3 \hat{\rho}_{\rm t}/4$ by considering that, for the two-electron spin state, there are 25\% and 75\% of spin singlet and triplet in the ensemble, respectively.
The density operators $\hat{\rho}_{\rm s}$ and $\hat{\rho}_{\rm t}$ for the spin singlet and the spin triplet have the form of  $\hat{\rho}_{\rm{s/t}} = \frac{1}{Z_{\rm{s/t}}} \sum_{i=0} \exp\left(-   E_{\rm{s/t}}^{(i)}  / (k_B T_{\rm{eff}} )  \right)  |i; {\rm{s/t}} \rangle \langle i; {\rm{s/t}}|$, where $E_{\rm{s/t}}^{(i)}$ and $|i; {\rm{s/t}} \rangle$ are the $i$-th eigenenergy and eigenstates of the spin singlet/triplet case obtained from the exact diagonalization, $\exp\left(\cdots\right)$ is the Boltzmann factor, and $Z_{\rm{s/t}}$ corresponds to the partition function. 

Using the thermal ensemble $\hat{\rho}$, we compute the transfer probabilities of the two electrons by using
\begin{equation}
P_{20} = \text{Tr} \left[ \hat{\rho} \int^{0}_{-\infty} dy_1 \, \int^{0}_{-\infty} dy_2 \, |y_1 \rangle \langle y_1 | \otimes |y_2 \rangle \langle y_2| \right],
\end{equation}
\begin{equation}
P_{02} = \text{Tr} \left[ \hat{\rho} \int^{\infty}_{0} dy_1 \, \int^{\infty}_{0} dy_2 \, |y_1 \rangle \langle y_1 | \otimes |y_2 \rangle \langle y_2| \right],
\end{equation}
\begin{equation}
P_{11} = 1 - P_{20} - P_{02},
\end{equation}
where $y_i$ is the transverse directional coordinate of electron $i=1,2$,
$y=0$ is reference location at the top of the tunnel barrier between the two QDs,
$|y_i \rangle \langle y_i|$ with $y_i > 0$ ($y_i < 0$) is the projector onto the states of electron $i$ in the upper (lower) QD.
We also compute the transfer probabilities $P_{10}$ and $P_{01}$ of a single electron by solving the Hamiltonian $(p_x^2 + p_y^2)/ (2m_e) + U(x,y)$.

Let us now reproduce the two-electron partitioning data shown in Fig.~\ref{fig:coldet}a of the main text.
Choosing an effective temperature of $k_B T_{\rm{eff}} = 1.42$~meV and a conversion factor of $\alpha = 0.095$~eV/V, we find that the computed  transfer probabilities $P_{20}$, $P_{11}$ and $P_{02}$ (with varying the detuning $\Delta$) in Fig.~\ref{sfig:ed}a well reproduce the experimental data.
These numerical results support the conclusion that Coulomb repulsion is the dominant mechanism in the observed antibunching.
We note that the chosen parameters are of the same order of magnitude with the effective thermal excitation energy $\varepsilon = 3$~meV and the conversion factor $\alpha=0.2$~eV/V estimated in Appendix~\ref{suppl:p11}. This implies that our equilibrium state ensemble with the effective temperature and the Boltzmann factor imitates well the non-adiabatic excitations in the non-equilibrium situation of our experiment. 

Furthermore, knowing that this model considers the exact confinement potential of the double-well potential, we can relate $P_{11}$ to the balance between the on-site Coulomb energy within one QD and the inter-dot Coulomb energy between the two moving QDs.
In particular, if the on-site energy is larger than the inter-dot energy, having one electron on each QD is thus the favourable state, which results in an increase of $P_{11}$. 

The dependencies of the calculated transfer probabilities on the barrier gate voltage $V_{\rm B}$ and the input power $P_{\rm in}$ shown in Fig.~\ref{sfig:ed}b and \ref{sfig:ed}c also provide deeper understanding of the experimental findings shown in Fig.~\ref{fig:tcw}a of the main text and in Fig.~\ref{sfig:saw}a, respectively. 
For instance, applying a more negative $V_{\rm B}$ -- which is equivalent to an increase of the distance $d$ between the two electrons -- reduces the inter-dot energy, resulting in an enhancement of $P_{11}$ (see Fig.~\ref{sfig:ed}b).
A stronger SAW amplitude -- controlled by the input power $P_{\rm in}$ -- has a similar effect.
In this case, the on-site Coulomb energy becomes larger in the modified potential profile, leading as well to an increase in $P_{11}$ (see Fig.~\ref{sfig:ed}c).

In summary, the good agreement between the results from the exact diagonalization method and the experimental data provides further evidence of the dominant role of Coulomb interaction in our system.

\section{Effective detuning dependence on barrier height}
\label{suppl:delta}

To investigate the effect of the barrier height in the TCW, we analyse the in-flight-partitioning data of two electrons that are sent simultaneously from the upper and lower source QDs.
Figure \ref{sfig:delta}a shows the effective detuning $\delta$ extracted from the partitioning data for three different barrier-gate voltages $V_{\rm B}$.
The red line shows a linear fit providing $\delta$ for the Bayesian model applied for Fig.~\ref{fig:tcw} in the main text.
Figure \ref{sfig:delta}b shows simulations of the maximum antibunching probability $P_{11}$ using the Bayesian model as a function of $V_{\rm B}$.
The data points from experiment are shown as reference.
Note that in the simulations $\sigma$ and $\Delta_{\rm S}$ are taken from single-electron partitioning measurements, so that $\delta$ is the only free parameter.
Assuming a constant $\delta$ (solid lines), the simulated results either over- or under-estimate $P_{11}$.
Using in the contrary the dependency extracted from a linear square fit from Fig.~\ref{sfig:delta}a (red line), the model shows a remarkable agreement over the whole voltage range.

\section{Antibunching dependency on effective thermal excitation}
\label{suppl:p11}

In the following we present a predictive investigation of the Coulomb-related antibunching rate by evaluating the Bayesian model assuming reduced excitation of the flying electrons.
Figure \ref{sfig:p11}a shows the maximum $P_{11}$ as a function of the single-electron partitioning width $\sigma$.
We express $\sigma$ as an effective thermal excitation $\varepsilon=\alpha \cdot \sigma$ where the gate alpha factor $\alpha=1/5$ is extracted from a fit by assuming an exponential distribution \cite{Takada2019}.
We find that by reducing the current excitation $\varepsilon$ by a factor of 3, the antibunching rate $P_{11}$ is beyond 99\%.
Comparing the simulated course of $P_{11}$ -- see Fig.~\ref{sfig:p11}b --, we expect a narrowing of the distribution for smaller excitation.
The saturation to 100\% represents the condition where the Coulomb-mediated antibunching is robust against small variations in the gate detuning $\Delta$.

\section{The role of SAW confinement}
\label{suppl:saw}

In the following we investigate the influence of SAW confinement amplitude on the antibunching process.
Figure \ref{sfig:saw}a shows the excess in antibunching probability $\Delta P_{11}$ extracted from the two-electron partitioning data for several applied input power $P_{\rm in}$ on the transducer.
We observe two regimes distinguished by a change in the slope around $P_{\rm in} \approx 24.5$~dBm.
Below this value, we know from the SAW amplitude calibration (see Fig.~\ref{sfig:sawampl}c) that the SAW confinement is not strong enough to avoid electron tunneling to subsequent minima.
For the region above the 95\% threshold for in-flight confinement \cite{Edlbauer2021}, $\Delta P_{11}$ gradually increases with SAW power.
A possible explanation is that, as $P_{\rm in}$ increases, the charging energy within each moving QD becomes larger, and thus overcoming the effective thermal excitation of the electrons.

To get a better understanding, we use the Bayesian model and extract the potential detuning $\delta$ as shown in Fig.~\ref{sfig:saw}b.
We observe that the course of $\delta$ is similar to $P_{\rm 11}$. 
From previous investigations, we know that $\delta$ also depends on the barrier height via $V_{\rm B}$.
While $V_{\rm B}$ controls mainly the coupling between the quantum rails that could affect the inter-dot energy -- we denote it as $U_{\rm inter}$ --, $P_{\rm in}$ changes the confinement potential within each moving QD, \textit{i.e.} on-site energy $U_{\rm site}$.
These results suggest that $\delta$ is a balance between $U_{\rm inter}$ and $U_{\rm site}$.

To check whether this hypothesis is valid, let us assume that the effective detuning is
\begin{equation}
\delta = \frac{|U_{\rm site} - U_{\rm inter}|}{\alpha}
\label{equ:col_delta}
\end{equation}
with $\alpha$ as the conversion factor from V to eV.

Let us first estimate the on-site energy $U_{\rm site}$. 
Since the confinement energy $\hbar \omega_x \approx 1$~meV along the SAW propagation direction ($x$) is smaller than the confinement energy $\hbar \omega_y \approx 3.6$~meV along the traversal direction ($y$) for the experimental conditions -- see Appendix~\ref{suppl:ed} --, we expect that the two-electron interaction is more sensitive to $\omega_x$ than $\omega_y$. 
Approximating the confinement along the $x$ direction as a parabolic potential (see Fig.~\ref{sfig:model}a) and ignoring its dependency on $\omega_y$, we can write $U_{\rm site}$ as
\begin{equation}
U_{\rm site} = \frac{1}{2} m_{e} \omega_x^2 (x_1^2 + x_2^2) + U_{\rm C}.
\label{equ:col_eintra}
\end{equation}
Here, $\beta=\frac{e^2}{4 \pi \varepsilon_r \varepsilon_0}\approx 111.6$~meV/nm is the Coulomb repulsion constant, $e$ is the elementary charge, $\varepsilon_0$ is the vacuum permitivity, $\varepsilon_r \approx 12.88$ is the dielectric constant of GaAs, $m_{e}\approx 0.067 m_0$ denotes for the electron effective mass in GaAs, $\omega_x$ is the parabolic confinement frequency, $x_i$ is the position of the electron $i$, and $U_{\rm C}$ corresponds to the unscreened Coulomb energy
\begin{equation}
 U_{\rm C} = \frac{\beta}{|x_1-x_2|}.
\label{equ:col_coulomb}
\end{equation}
Expressing Eq.~\ref{equ:col_eintra} in terms of the on-site separation $\Delta l_{\rm site} = |x_1 - x_2|$, we find via $\delta U_{\rm site}/\delta \Delta l_{\rm site} = 0$ that the lowest energy of the system is
\begin{equation}
U_{\rm site} = \frac{3}{2^{\frac{4}{3}}} \left (m_e \omega_x^{2}{\beta}^2 \right )^{\frac{1}{3}}
\end{equation}
at the optimum distance of
\begin{equation}
\Delta l_{\rm site} = \left ( \frac{2 \beta}{m_{e} \omega_x^2} \right ) ^{\frac{1}{3}}.
\end{equation}
If only few electrons are present, the ground state of the system found via this classical approach is equivalent to solving the quantum Hamiltonian \cite{Ciftja2009}.

Substituting $\omega_x$ from Eq.~\ref{equ:omega_x_P} in Eq.~\ref{equ:col_eintra}, we reach to the final relation
\begin{equation}
U_{\rm site}(P_{\rm in}) = \frac{3}{2} \left ( \frac{\pi \beta}{\lambda} \right )^{\frac{2}{3}} \left ( A_0 10^{\frac{P_{\rm in}}{20}} \right )^{\frac{1}{3}}.
\label{equ:col_eintra_p}
\end{equation}
For the maximum applied power $P_{\rm in}=28$~dBm, two electrons occupying the same moving QD would be separated by $\Delta l_{\rm site} \approx 65$~nm which results in $U_{\rm site} \approx 2.6$~meV. 

Let us now estimate $U_{\rm inter}$ via the inter-dot distance $\Delta l_{\rm inter}$ as depicted in Fig.~\ref{sfig:model}b.
In the limiting case of unscreened Coulomb repulsion, the energy is simply 
\begin{equation}
U_{\rm inter} = \frac{\beta}{\Delta l_{\rm inter}}.
\end{equation}
Since the co-propagating electrons have a separation $\Delta l_{\rm inter} \approx 230$~nm, we estimate $U_{\rm inter} \approx 0.5$~meV.
Note that owing to $U_{\rm site} > U_{\rm inter}$, the electron pair tends to occupy different moving QDs, and hence the observation of the Coulomb-induced antibunching effect.

Having expressed $\delta$ as a function of the input SAW power $P_{\rm in}$, we employ the equations~\ref{equ:col_delta} and~\ref{equ:col_eintra_p} to reproduce the experimental data shown in Fig.~\ref{sfig:saw}b where $\alpha$ is the only fitting parameter.
Using $\alpha \approx 1/9$ eV/V, our estimation (red line) shows a good agreement with the extracted detuning $\delta(P_{\rm in})$.
These results confirm our expectation that the SAW amplitude modifies the on-site energy, and thus the antibunching probability.

\bibliography{references.bib}

\renewcommand{\thefigure}{S\arabic{figure}}    
\setcounter{figure}{0} 

\begin{figure*}[t]
\includegraphics[width=\dcwidth]{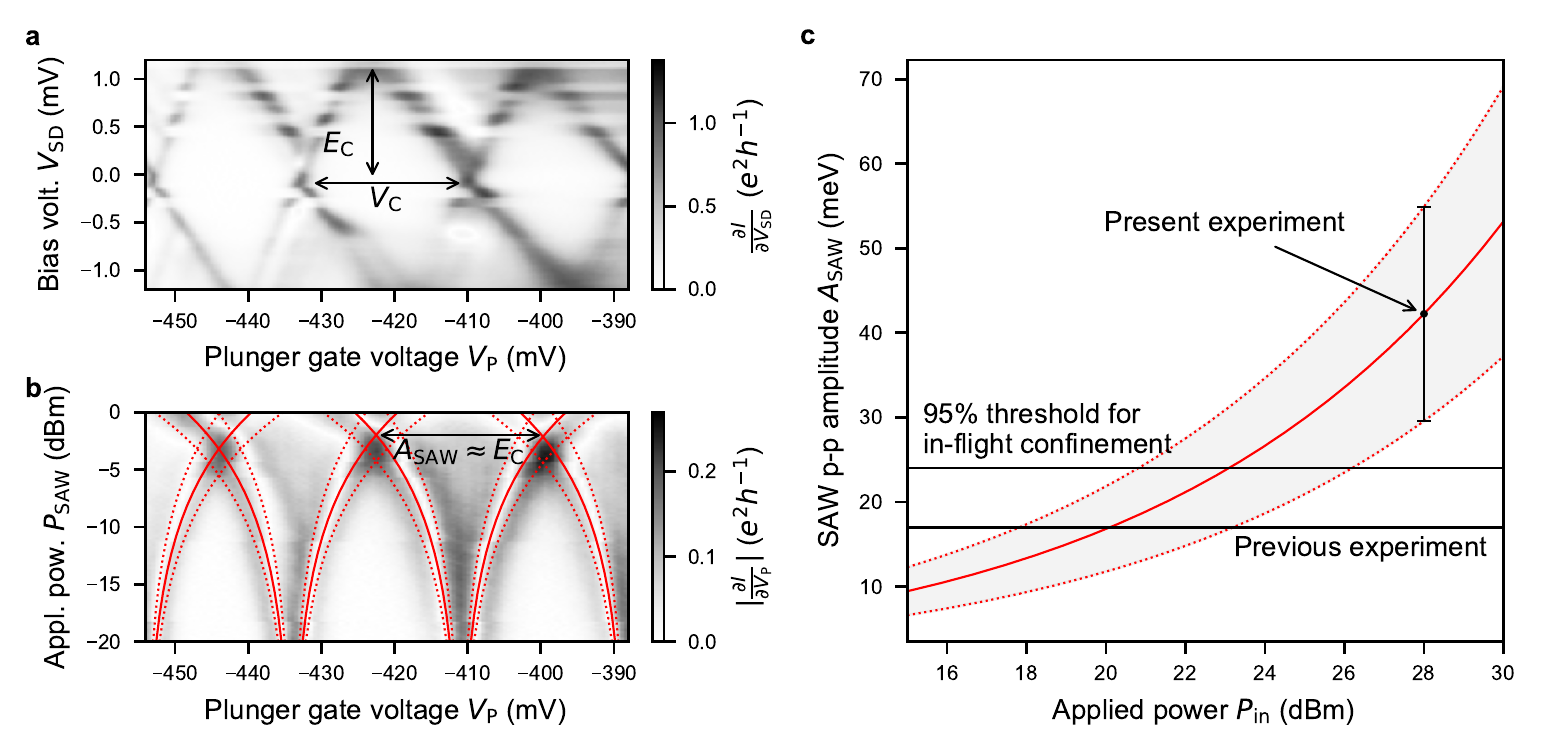}
\caption{Extrapolation of SAW amplitude from quantum-dot modulation.
    \textsf{\textbf{(a)}}
    Coulomb diamonds.
    The data shows a transconductance measurement as function of the bias voltage $V_{\rm SD}$ and the voltage $V_{\rm P}$ applied on the plunger gate.
    The arrows indicate the charging energy $U_{\rm C}$ and the gate-voltage period, $V_{\rm C}$, of the resonances. 
    \textsf{\textbf{(b)}}
    Broadening of the Coulomb peaks as function of $P_{\rm in}$.
    The double-headed arrow indicates the value of $P_{\rm in}$ where the peak-to-peak amplitude $A_{\rm SAW}$ of the SAW matches $U_{\rm C}$.
    \textsf{\textbf{(c)}}
    Extrapolation of $A_{\rm SAW}$ for $P_{\rm in}$ up to 30 dBm.
    The errorbar indicates the estimation of $A_{\rm SAW}$ for the present experiment.
    Horizontal lines serve as reference SAW amplitudes from Ref. \cite{Takada2019} (lower) and \cite{Edlbauer2021} (upper).
    \label{sfig:sawampl}
}
\end{figure*}

\begin{figure*}[t]
\includegraphics[width=\dcwidth]{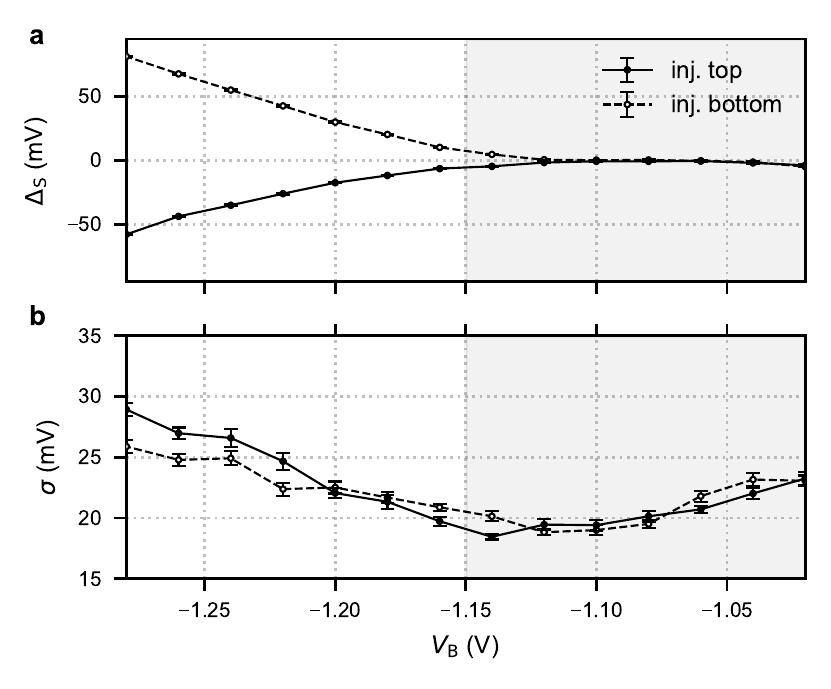}
\caption{Barrier dependence of in-flight partitioning.
    Influence of the barrier gate voltage $V_{\rm B}$ on \textsf{\textbf{(a)}} the offset $\Delta_{\rm{S}}$ and \textsf{\textbf{(b)}} the width $\sigma$ of the in-flight-partitioning data measured for SAW-driven transport of a single electron from the upper (point; solid line) and lower (circle; dashed line) source QD.
    The shaded region highlights the voltage range where $\Delta_{\rm{S}}$ is similar for both injections.
    \label{sfig:dircoup}
}
\end{figure*}

\begin{figure*}[t]
\includegraphics[width=\dcwidth]{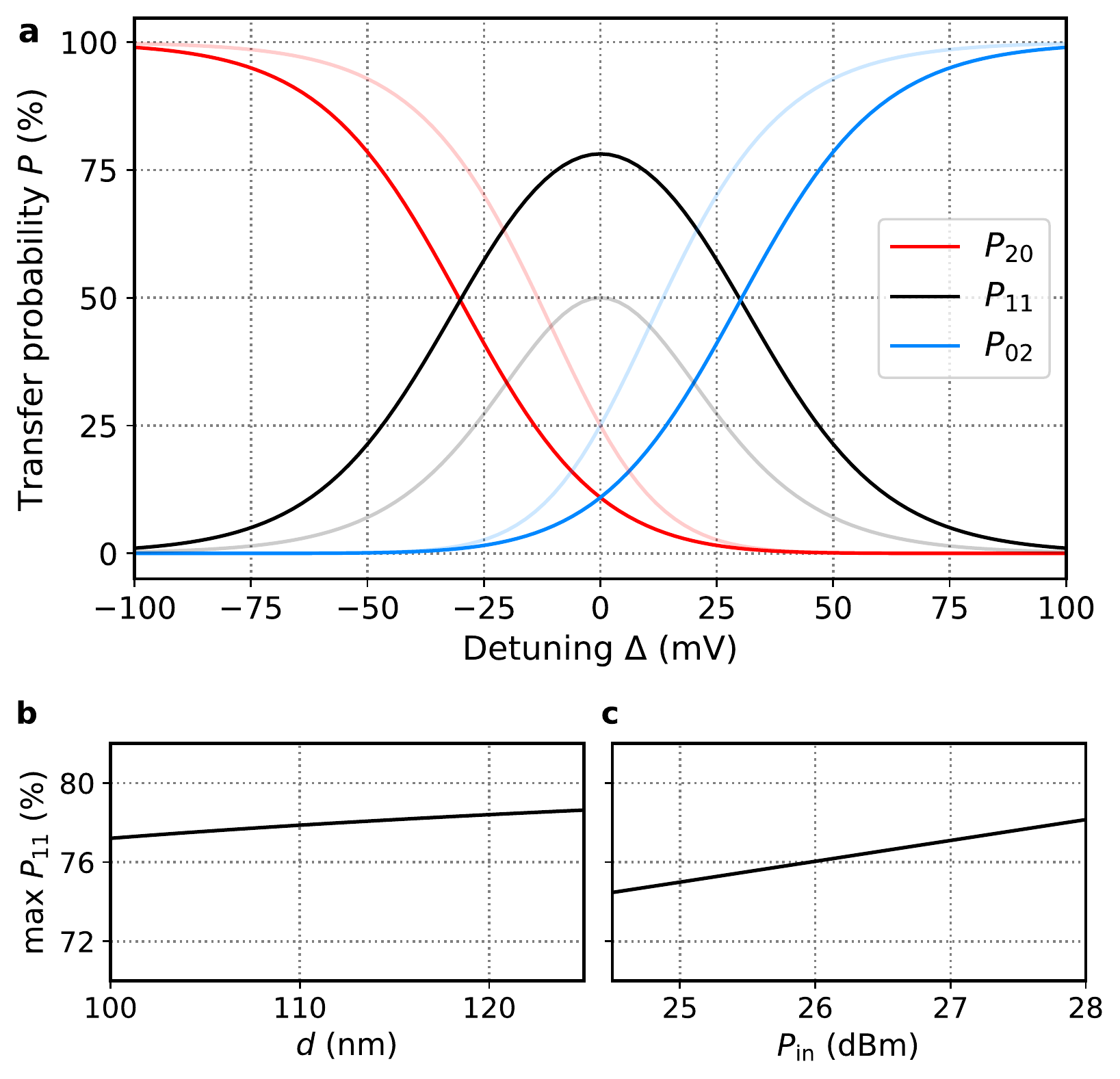}
\caption{Transfer probabilities of a thermal ensemble obtained by exact diagonalization.
    \textsf{\textbf{(a)}}
    Probabilities $P_{20}$, $P_{11}$, and $P_{02}$ with Coulomb interaction (solid lines) as a function of the detuning $\Delta$. 
    The non-interacting cases are shown as $P_{10}^2$ (semi-transparent red),  $2P_{10} P_{01}$  (semi-transparent black), $P_{01}^2$  (semi-transparent blue), which correspond to the transfer probabilities of an electron pair in different potential minima accompanying the SAW train. 
    The calculations are performed with a SAW input power $P_{\rm in} = 28$~dBm and a distance $2d = 230$~nm between the moving double QDs. 
    \textsf{\textbf{(b)}}
    Maximum value of $P_{11}$ (at $\Delta = 0$) as a function of $d$ with $P_{\rm in} = 28$~dBm.
    \textsf{\textbf{(c)}}
    Maximum $P_{11}$ as a function of $P_{\rm in}$ for $2d = 230$~nm.
    \label{sfig:ed}
}
\end{figure*}

\begin{figure*}[t]
\includegraphics[width=\dcwidth]{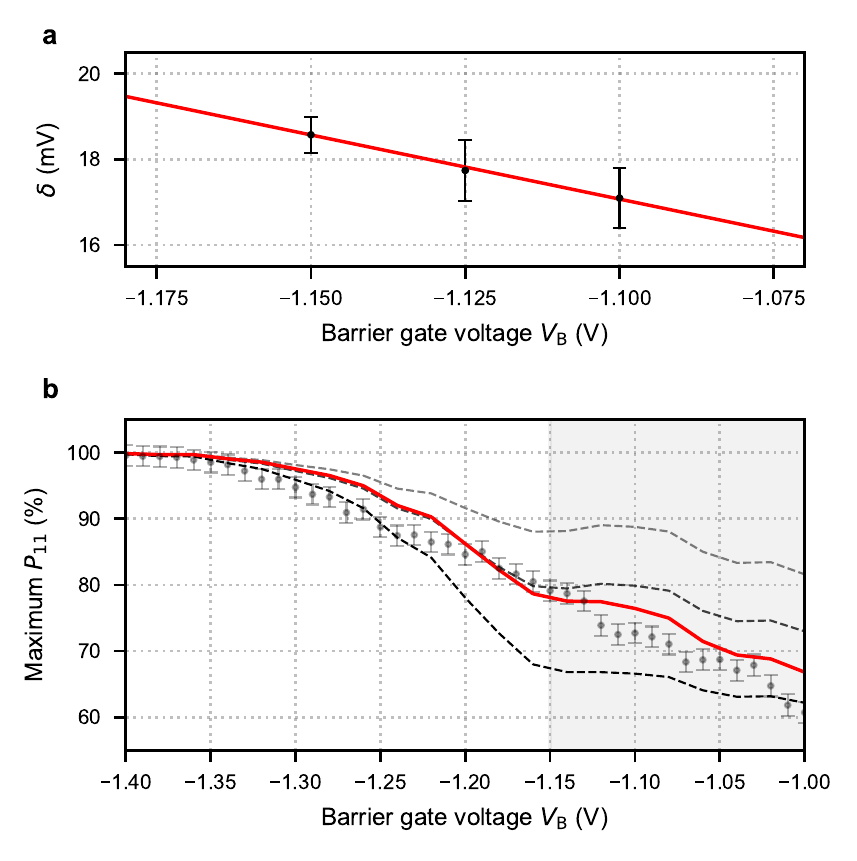}
\caption{Extraction of effective detuning.
    \textsf{\textbf{(a)}}
    Effective gate detuning $\delta$ extracted from a fit with Bayesian model of two-electron-partitioning data for different barrier voltages $V_{\rm B}$.
    The data points follows a linear function with slope $\approx -30$~mV/V.
    \textsf{\textbf{(b)}}
    Simulated traces of the maximum antibunching probability $P_{11}$ with experimental data points as reference.
    Dashed lines are obtained considering constant $\delta \in [10, 20, 30]$ mV (from black to grey).
    The red solid line shows the expected course by using the $\delta$-dependency extracted from (a).
    \label{sfig:delta}
}
\end{figure*}

\begin{figure*}[t]
\includegraphics[width=\dcwidth]{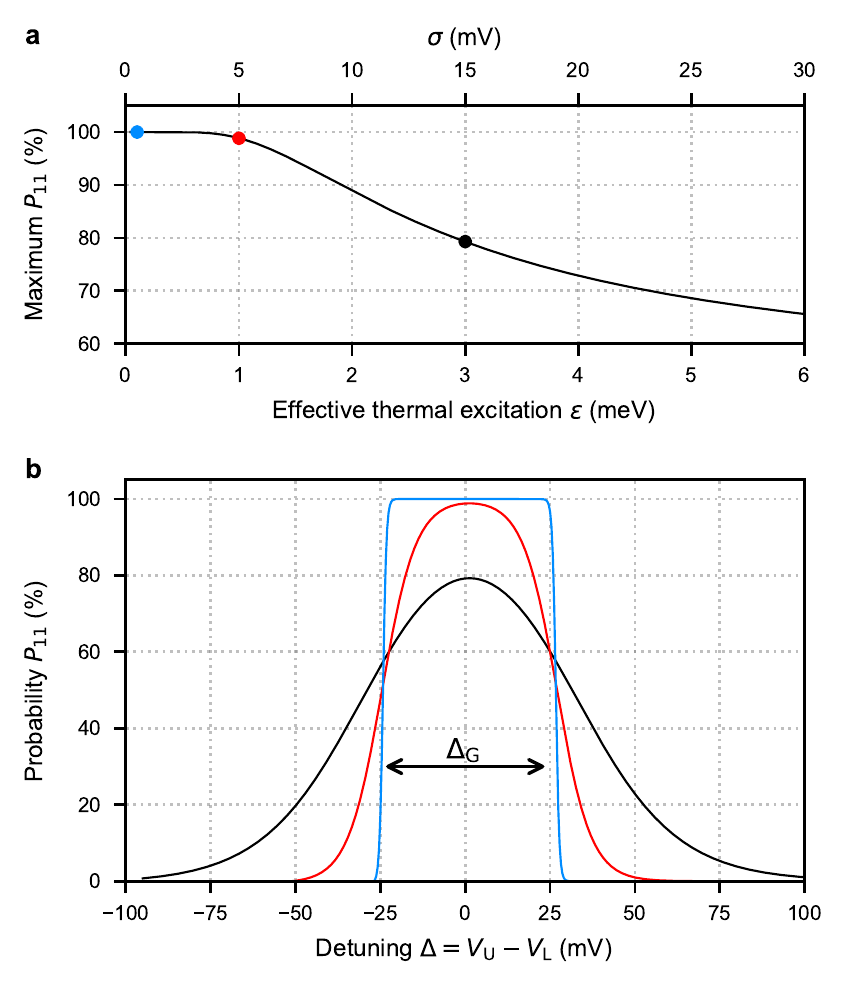}
\caption{Effective thermal excitation.
    \textsf{\textbf{(a)}}
    Simulations based on the Bayesian model of the maximum antibunching probability $P_{11}$ as a function of the effective thermal excitation $\varepsilon$ of both electrons. 
    Here, we employed the induced detuning $\delta=18.5$~mV extracted for $V_{\rm B}=-1.15$~V.
    For the current experimental condition ($\varepsilon=3$~meV; black point), $P_{\rm 11}\approx80$\%.
    The threshold of $P_{11} = 99$\% is reached at $\varepsilon=1$~meV (red point).
    \textsf{\textbf{(b)}}
    Evolution of $P_{11}$ as a function of $\Delta$ for different $\varepsilon \in [0.1, 1.0, 3.0]$~meV (blue, red and black).
    When the electron U and L are in the ground state, $P_{\rm 11}$ is saturated at 100\% for a detuning width $\Delta_{\rm G}=2\delta+|(\Delta_{\rm S}^{\rm U}-\Delta_{\rm S}^{\rm L})|$ (double-headed arrow).
    \label{sfig:p11}
}
\end{figure*}

\begin{figure*}[t]
\includegraphics[width=\dcwidth]{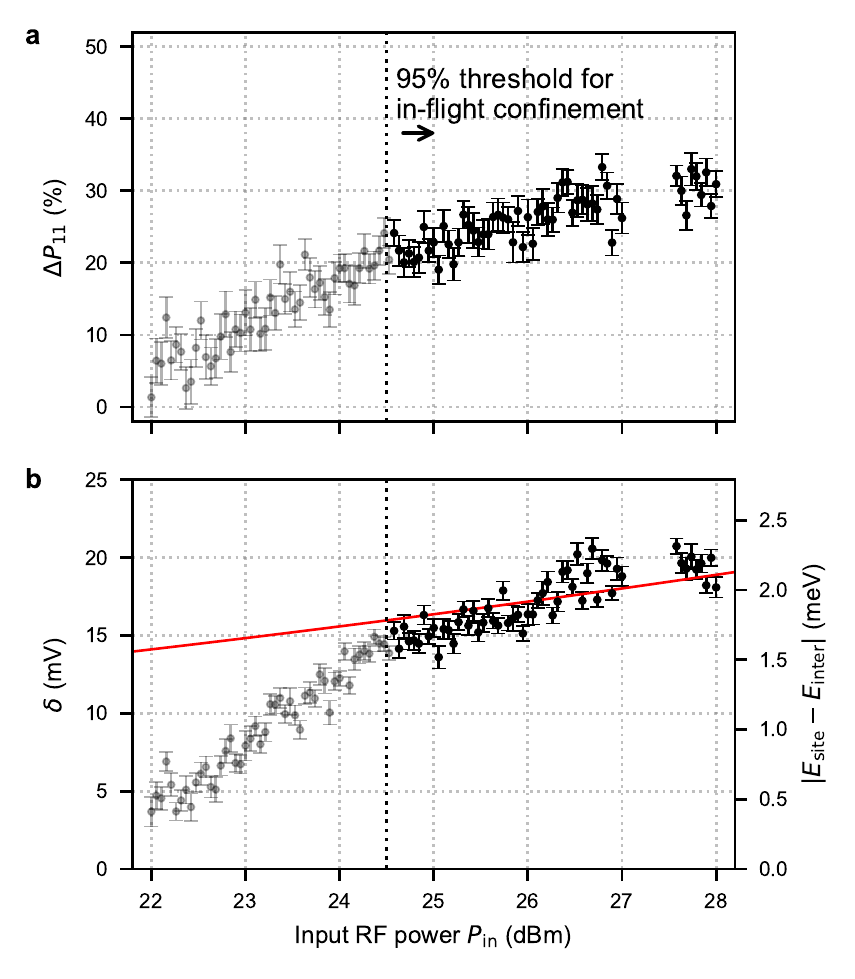}
\caption{SAW confinement.
    \textsf{\textbf{(a)}}
    Excess of antibunching probability $\Delta P_{11}$ subtracted from two-electron-partitioning data with and without interaction.
    The missing data points around 27~dBm are due to technical problems during the measurements.
    \textsf{\textbf{(b)}}
    Extracted $\delta$ from a fit using the Bayesian model of two-electron-partitioning data at each applied power.
    The expected evolution (red line) above the 95\% confinement threshold is calculated by approximating the SAW potential to a parabolic QD.
    \label{sfig:saw}
}
\end{figure*}

\begin{figure*}[t]
\includegraphics[width=\dcwidth]{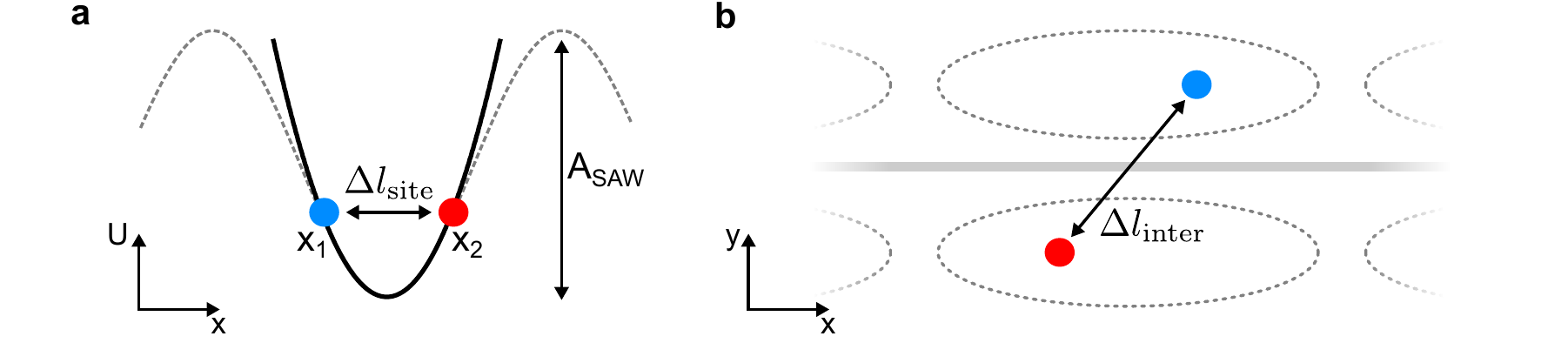}
\caption{Schematic of electron-pair locations.
    \textsf{\textbf{(a)}}
    SAW confinement potential (dashed grey line) along the propagation direction $x$ with peak-to-peak amplitude $A_{\rm SAW}$.
    The parabolic approximation is depicted on top (black line).
    The electrons (blue and red) at positions $x_1$ and $x_2$ are separated by $\Delta l_{\rm site}=|x_2-x_1|$.
    \textsf{\textbf{(b)}}
    Schematic top view in the coupling region where the pair of electrons are located at different moving QDs (dashed circles).
    The inter-dot separation $\Delta l_{\rm inter}$ is indicated by the double-headed arrow.
    \label{sfig:model}
}
\end{figure*}

\end{document}